ORIGIN OF RADIO-QUIET CORONAL MASS EJECTIONS IN FLARE STARS

D. J. Mullan and R. R. Paudel

Department of Physics and Astronomy, University of Delaware, Newark DE 19716

ABSTRACT

Type II radio bursts are observed in the Sun in association with many coronal mass ejections (CME's). In view of this association, there has been an expectation that, by scaling from solar flares to the flares which are observed on M dwarfs, radio emission analogous to solar Type II bursts should be detectable in association with M dwarf flares. However, several surveys have revealed that this expectation does not seem to be fulfilled. Here we hypothesize that the presence of larger global field strengths in low-mass stars, suggested by recent magneto-convective modeling, gives rise to such large Alfven speeds in the corona that it becomes difficult to satisfy the conditions for the generation of Type II radio bursts. As a result, CME's propagating in the corona/wind of a flare stars are expected to be "radio-quiet" as regards Type II bursts. In view of this, we suggest that, in the context of Type II bursts, scaling from solar to stellar flares is of limited effectiveness.

1. **INTRODUCTION**

For several decades, it has been believed that flares on low-mass stars are due to magnetic processes analogous to those which are observed in solar flares (e.g. Gershberg 1970; Mullan 1976). In view of this, it might seem worthwhile to perform scalings of various parameters from solar flares to those of stellar flares. In some cases, such scalings work out quite well, e.g. in determining the basic parameters of flare plasma (Schmitt et al 1989), and in understanding the frequency distributions of flare energies (Mullan and Paudel 2018).

However, some recent works suggest that scaling from solar to stellar flares may not always lead to satisfactory conclusions. For example, the Sun is known to generate coronal mass ejections (CME's) in association with many (but by no means all) flares. Specifically, during the years 1996-2010, coronagraph data permitted the detection of 12,433 CME's while X-rays were detected from 22,688 flares (Youssef 2012), i.e. CME's were detected in 50-60% of solar X-ray flares. Thus, the data indicate that not all X-ray flares in the Sun are accompanied by detectable CME's. The correlation between CME's and X-ray flares was found to be best when the Sun is at a high activity level, possibly because in larger flares, the CME's are observed to be faster and wider and therefore easier to detect. Aarnio et al (2011) identified almost 1000 solar flares from the years 1996-2006 for which the necessary parameters could be evaluated, and found a statistically significant correlation between the mass of the CME and the X-ray flux. In view of these results, we might expect that in considering flares on M dwarf stars, where flare energies can exceed the largest solar flare energies by factors of up to 100-1000, there should be a strong correlation between the physical parameters of flare events and those of CME's.

However, the expectation has not turned out so well when observers have searched for CME's using two distinct observational approaches: Balmer line profiles, and Type II radio observations.

**First, in the context of Balmer lines, Leitzinger et al (2014) searched for flares and flare-related CME's in 28 dK-dM stars in a young cluster. Their technique was based on the following statement in their paper: "Pronounced Balmer line asymmetries are likely to be related to CME's as no other sporadic phenomenon is known to produce velocities of several hundred km/sec".** They cited Houdebine et al (1990) who had reported observations of a large flare on AD Leo in the H$\gamma$ line where a blue-shifted feature was observed at 5800 km/sec with an estimated mass of $8 \times 10^{17}$ gm. Leitzinger et al also cited other previous searches of Balmer lines, none of which had produced any convincing evidence of a CME. But in view of the Houdebine et al success with H$\gamma$, Leitzinger et al (2014) obtained spectra in the vicinity of another Balmer line (H$\alpha$) and searched for a Doppler-shifted feature in the blue wing of the line profile. Their spectral resolution was such that they could detect CME's only if the projected velocity exceeded their resolution limit, i.e. 135 km/sec. Moreover, in view of their finite S/N ratios, they could detect a CME with a confidence of >3$\sigma$ against the noise background only if the energy flux of the CME in H$\alpha$ exceeded $1.4 \times 10^{-16}$ ergs/cm$^2$/sec. In 4 of their target stars, they reported seeing features which were blue-shifted with speeds ranging from ~175 to ~300 km/sec: however, none of the features was large enough to satisfy the >3$\sigma$ criterion. They admitted that since CME's are ejected from a star at arbitrary angles, observers can measure only the projected velocity relative to the line of sight. This led to their admission that "the identification and interpretation of CME signatures" is "challenging". Although Leitzinger et al (2014) certainly did observe H$\alpha$ flares in their target stars, their conclusion regarding CME's was quite definite: "We did not detect any clear signs of stellar CME's".

Using the same observational approach as Leitzinger et al, Vida et al (2016) subsequently undertook a search for CME's in other flare stars. Vida et al reported that the detection rate of CME's in H$\alpha$ was "much lower than expected from extrapolations of the solar flare-CME relation". Moreover, Korhonen et al (2017) also used the H$\alpha$ line **in the same way that Leitzinger et al (2014) had done** to observe G, K, and M stars in 5 clusters of various ages: they did detect H$\alpha$ flares, but they detected no CME's. Their report includes the remark that their failure to detect CME's might indicate that "there actually are only few of them [CME's]". **In a very recent study along similar lines, Vida et al (2019) have examined an even larger data set (5500 spectra) for Balmer line asymmetries: they find that "in most cases, the detected speed does not reach the surface escape velocity".**

Second, in the context of radio observations, Crosley and Osten (2018) searched for Type II radio bursts in a 64-hour long observing period aimed at the active flare star EQ Peg, but detected none. They concluded that this "casts serious doubt on the assumption that a high flaring rate [in an M dwarf] corresponds to a high rate of CMEs". Moreover, no evidence for Type II bursts was obtained by Villadsen and Hallinan (2018) in 58 hours of observing 5 active M dwarfs: those authors pointed out that Type II bursts in the Sun are "often driven by super-Alfvenic CME's", suggesting the possibility that in flare stars, CME's moving faster than the Alfven speed may be rare. We note that Aarnio et al (2012), in their application of the solar correlation between CME mass and X-ray flare energy to pre-main sequence stars, made no mention of Type II bursts. As a result, it is unknown whether or not the CME's discussed by Aarnio et al in pre-main sequence stars were accompanied by Type II radio bursts: as such, the work of Aarnio et al (2012), although obviously of interest in the general context of stellar flares, will not be discussed any further in the present paper.

The Hα data (Leitzinger et al) and the radio data (Crosley & Osten; Villadsen & Hallinan) suggest that simple scalings from the solar case do not yield reliable predictions about CME's in flare stars.

One possible explanation for the scarcity of CME's in flare stars has recently been pointed out in the context of the flare star Trappist-1 (Mullan et al 2018). In an MHD modeling study of solar CME's, Alvarada-Gomez et al (2018) have found that if the global **field is strong enough (of order 75 G), CME's with kinetic energy lower than a certain value can in fact be suppressed: it is as if the "strapping force" of overlying field lines (see Inoue et al 2018) may be so strong that the kinetic energy of the CME is not sufficient to allow "break out" from the low corona. Or it may be that if the CME is a rising filament containing a magnetic flux-tube, the relative orientations of the strapping field lines and of the filament field lines are not conducive to fast reconnection: in such cases, the filament may not be able to find a way to escape from the Sun, thereby confining the CME to the low corona. As a result, the CME might never reach a coronal location where the CME speed exceeds the MHD fast-mode speed (see Section 2 below): if this happens, no coronal shock would be generated, thereby leading to no Type II radio emission.** Vida et al (2019) also state that their negative Balmer-line searches "may support the idea" of CME suppression by strong coronal fields.

In the case of the flare stars GJ65A/B (MacDonald et al 2018) and Trappist-1 (Mullan et al 2018), magneto-convective modeling has led to estimates of the global magnetic field strengths in these stars. In both systems, the global field is found to exceed 1 kG in strength: this is considerably in excess of the critical fields of 75 G which Alvarado-Gomez et al (2018) found capable of suppressing CME's in the Sun. In view of this, Mullan et al (2018) have suggested that CME's in Trappist-1 could be suppressed even if the kinetic energies are ~100 times larger than the energy of the largest flare observed (so far) in Trappist-1.

In the present paper, we consider another effect of the strong magnetic fields which, according to magneto-convective modeling, may be operative in flare star coronae. Specifically, we attempt to quantify the suggestion of Villadsen and Hallinan (2018) concerning the lack of super-Alfvenic phenomena.

2. **TYPE II RADIO BURSTS AND THE ALFVEN SPEED**

According to Wild et al (1963), the "great majority" of solar flares are so small that the only radio emissions they exhibit are those which are categorized as Types III and V at meter-wavelengths. But in "certain large flares", further radio phenomena (often highly energetic) develop. Among these phenomena, Type II radio bursts in the Sun are observed at meter-wavelengths, and are characterized by slowly drifting from high to low frequencies. The frequency drifts are observed to occur at rates of order 1 MHz/sec, and the bursts last for time scales of 5-30 minutes. **At the time of the Wild et al (1963) paper, reliable observational information about the physical parameters of CME's was lacking: spacecraft images of CMEs did not become available until the early 1970's (Tousey 1973; MacQueen** et al 1974). As a result, the relationship between Type II bursts and CME's (the principal topic of the present paper) could not be addressed by Wild et al (1963).

In a study of meter-wave Type II bursts during a time when CME information was simultaneously available (1996-2008, during the declining phase of solar cycle 23), Gopalswamy et al (2009) reported specifically on the relation between Type II bursts and CME's. Of particular importance, they identified that Type II bursts did not begin until the CME reached a certain height in the corona: the heliocentric distance at which CME-driven shocks form was determined to be ~1.5 $R$ . Gopalswamy et al noted that this is the radial location where the Alfven speed in the corona is believed to have a *minimum* value. The existence of a minimum in $V_A$ at a certain radial location arises because the coronal field has two principal components: (i) strong fields associated with active regions, and (ii) the global (dipolar) field of the Sun. Field (i) is strongest near the Sun, and creates $V_A$ which falls off rapidly with increasing height. Field (ii) creates $V_A$ with a small value close to the Sun, where the density is largest. The superposition of the radial behavior of the $V_A$ components (i) and (ii) inevitably leads to a minimum in the $V_A$ profile at a certain radial location, of order $1.5R_*$ . (See e.g. Mann et al 2003, their Fig. 6.)

Moreover, at the radial location where the Alfven speed in the solar corona/wind has a maximum value (~3-4 $R$ ), Gopalswamy et al found that CME's of moderate speed produced either no shock at all (i.e. no Type II burst), or only a weak shock. These are significant conclusions for the present paper because they indicate the important role played by the coronal Alfven speed in CME-driven shock formation, which in turn is related to Type II generation. As regards the variation of CME's during the solar cycle, Gopalswamy et al (2009) reported that the number of CME's per solar rotation was observed to be 10 or more over an interval of several years surrounding the solar maximum in 2001-2002: the peak number of events in a single solar rotation was 19. Near solar minimum, the number of Type II meter-wave bursts per rotation fell to only 1, or in some cases, zero. This suggests that if we wish to observe CME's in flare stars, the chances for detection will be optimized if we observe stars with high levels of activity.

The speed $V_{II}$ with which the Type II source moves through the corona is observed to be of order $10^3$ km/sec (Wild et al 1963). The empirical values of $V_{II}$ are considerably larger than the isothermal speed of sound $V_s$ in the coronal protons: at temperatures which are typical of the solar corona, i.e. $T \approx 1-2$ MK, the numerical value of $V_s$ is of order 100-140 km/sec. The fact that $V_{II}$ exceeds $V_s$ by a factor of order 10 leads to the conclusion that Type II bursts are not hydrodynamic in nature. Instead, a Type II burst is believed to be associated with a magnetohydrodynamic (MHD) disturbance ascending upwards through the corona (Wild et al 1963). When the MHD disturbance reaches an altitude where the local physical conditions are favorable for the formation of an MHD shock, plasma radiation can be excited (Zaitsev 1969): this can then be converted into electromagnetic radiation via scattering off inhomogeneities in the ambient corona. The electromagnetic radiation associated with the MHD disturbance moving upwards into plasma with lower density gives rise to a Type II burst drifting towards lower frequencies in solar radio data. If the radial profile of density in the corona is known, the rate of frequency drift can be converted to a propagation speed $V_{II}$ .

An important conclusion of Zaitsev (1969) is that Type II bursts will be generated only if the MHD shock speed $V_{II}$ is large enough that the Alfvenic Mach number, $M_A = V_{II}/V_A$ , exceeds a critical value of 1.1-1.2. Given a value for $V_{II}$ , it is clearly easiest to satisfy this criterion at the spatial location where

$V_A$ has its smallest numerical value, i.e. in the vicinity of r = 1.5$R_*$. This can help to explain why Gopalswamy et al (2009) reported that Type II shocks do not begin in the corona until a the MHD shock reaches a radial location of order 1.5$R_*$. It also helps to explain another conclusion of Gopalswamy et al (2009), namely, CME's with moderate speeds do not generate Type II bursts at radial distances of 3-4$R_*$: at such distances in the solar corona, $V_A$ reaches its maximum value, of order 700-800 km/sec (see Mann et al 2003: their Fig. 6). With such large local values of $V_A$, it becomes more difficult for a CME with a moderate speed (e.g. < 700 km/sec) to drive the local value of $M_A$ above the requisite critical value of 1.1-1.2. As a result, it is not surprising that such CME's may not generate Type II bursts at all at radial locations of 3-4$R_*$.

Independent evidence for the presence of an excitation agent spreading out from a solar flare at speeds of order $10^3$ km/sec is provided by Hα waves reported by Moreton (1960) and by coronal waves observed in EUV (Klassen et al 2000). The MHD nature of Moreton waves was confirmed by Uchida (1968) in his study of the propagation of the waves in the vicinity of active regions: the waves were observed to avoid regions of high Alfven speed. Such avoidance behavior (as well as focusing of fast-mode waves into regions of low Alfven speed in the solar wind [Mullan et al 2003]) occurs because of the laws of refraction which are obeyed by a fast mode MHD wave moving through an inhomogeneous medium.

Provided that the Alfven speed $V_a$ is much less than the speed of light, the general formula for the speed $V(MHD)$ of a fast mode MHD wave is $V(MHD)^2 = V_a^2 + V_s^2$ (Spitzer 1962, p. 64). In coronal conditions, where $V_s$ is in general small compared to the Alfven speed $V_a$, the propagation speed $V(MHD)$ is essentially equal to $V_a$. Therefore, the general condition for a disturbance in the corona to create an MHD shock is essentially that the speed of the disturbance must exceed $V_a$.

However, we note that, despite the closeness in magnitude of the speed of the fast-mode $V(MHD)$ in the corona to the speed of the Alfven mode $V_a$, there is a fundamental difference in the way in which these two modes propagate as a function of the angular direction relative to the field (Spitzer 1962, p. 67). The Alfven mode propagates fastest (with speed $V_a$) when the wave is moving parallel to the field, but slowest (speed = 0) when the wave attempts to move perpendicular to the field. On the other hand, the fast-mode wave always moves at least as fast as $V_a$, and reaches its fastest speed when the wave moves perpendicular to the field: however, in coronal conditions, where $V_s/V_a \approx 0.1$, even the fastest speed of the fast-mode wave differs from $V_a$ by no more than a percent or so. Thus, fast-mode waves in the corona have the advantage of a rather well defined speed ($V_a$) which remains more or less the same for waves propagating in arbitrary directions relative to the direction of the field. That is, fast mode waves do not suffer from the restrictions imposed on Alfven waves as regards the directions in which they can propagate. Because of the differences between Alfven (non-compressive) waves and fast-mode (compressive) waves, the Alfven waves are not subject to the same refraction laws as fast-mode waves (Uchida 1968): as a result, when fast-mode waves are refracted away regions where the Alfven speed is high, the Alfvenic waves are observed to become dominant in those regions (Smith et al 2001).

The Alfven Mach number of a disturbance moving at speed V is defined as $M_A = V/V_a$. In order to generate Type II bursts in the Sun, Smerd et al (1975) found that their observational data on radio bursts at meter wavelengths (i.e. at frequencies of 50-200 MHz) required $M_A$ to exceed a value of about 1.2. Smerd et al noted that this empirical lower limit on $M_A$ is consistent with theoretical work (Zaitsev 1969) which shows that the generation of plasma radiation requires the shock to have $M_A$ in excess of 1.1-1.3 at altitudes of 1-2$R$ in the corona. These critical values of $M_A$ are sufficiently close to unity that we will make no significant error if we consider that the onset of Type II bursts can be expected when an MHD disturbance propagates in the corona at a speed in excess of a critical value: and that critical value that is essentially equal to $V_a$.

## 3. RADIAL PROFILE OF THE ALFVEN SPEED IN THE CORONA/WIND

In non-relativistic conditions, $V_a$ is given by the usual formula $B/\sqrt{(4\pi n m_p)}$. Here, $B$ is the (total) field strength, $n$ is the number density of protons, and $m_p$ is the proton mass. However, in regions of strong fields and/or low densities, the usual formula must be replaced by $V_a = c/K^{0.5}$ where $K = \sqrt{(1 + 4\pi n m_p c^2/B^2)}$ where c is the speed of light (Spitzer 1962, p. 62). This formula ensures that $V_a$ never exceeds the speed of light.

To obtain the radial profile $V_a(r)$, we need to know the profiles of $n(r)$ and $B(r)$. We now discuss these in turn.

### 3.1. The density profile

Mann et al (1999) have suggested that $n(r)$ can be obtained for an isothermal corona using an algebraic solution for wind speed $v(r)$ (their eq. (9)). Then the constancy of the mass loss rate allows $n(r)$ to be calculated from $r^2 n(r) v(r) = C$ where $C$ is related to the mass loss rate by $dM/dt = 4\pi m_p C$. Using solar wind data at 1 AU, Mann et al find that for the solar mass loss rate ((dM/dt) = 2 x $10^{-14}$ solar masses per year), the constant $C$ has the value 6.3x $10^{34}$ per second.

In order to consider flare stars, we need to ask: what would be an appropriate value for $C$ in the case of such objects? The answer depends on the mass loss rate dM/dt from these stars. Wood (2018) has summarized the values obtained for dM/dt for a number of systems for which the profile of the Lyman-α have been obtained using the Hubble telescope. Theory predicts that a "hydrogen wall" should build up ahead of the astrosphere surrounding a mass-losing star as the star moves through the interstellar medium. Wood's sample includes G, K, and M stars, some in binaries, others believed to be single. Since we are interested in flare stars here, we confine attention to K and M stars in Wood's sample: for those 14 stars, Wood lists (dM/dt)* values which range from 0.059 to 55.7 times (dM/dt) with a median value of (dM/dt)* between 1.54 and 4 times (dM/dt). The mass loss rates in binary stars may be enhanced by tidal forces which give rise to stronger fields than those in single stars. If we confine attention to the 8 single stars in Wood's sample,for those stars, the median value of (dM/dt)* is found to be 0.5 times (dM/dt).

These results suggest that when we consider the winds of flare stars and evaluate the $n(r)$ profile, we will typically not make errors in $n(r)$ of more than roughly 2 (either above unity or below unity) if we use the solar value of the constant $C$. The results to be presented below are obtained by assuming the solar value of $C$.

For the sake of completeness, we summarize here how we determined the velocity profile *v(r)*. The goal is to solve for *v* as a function of *r* for all values of *r* from close to the stellar surface ($r \approx r_*$) out to a radial distance of $200r_*$. (In the Sun's case, the latter radial location lies almost at the Earth's orbit.) To achieve this goal, we note that an equation for the wind speed was derived by Parker (1963: his eqn 5.43) for the case of an isothermal corona with temperature *T*. In such a corona, there exists a critical velocity $v_c$ which is defined by $v_c = \sqrt{kT/\mu m_p}$, i.e. the isothermal sound speed in a medium with mean molecular weight μ. (Here, *k* is Boltzmann's constant, and $m_p$ is the proton mass.) There also exists a critical radius $r_c = GM_*/2v_c^2$. Parker plotted solutions for *v(r)* (see his Fig. 6.1) for a range of *T* values from 0.5 to 4 MK. A convenient formulation of the isothermal *v(r)* equation has been given by Mann et al (1999): their eq. (9) can be written as

y – ln(y) = 4ln(x) + 4x -3     (eq. 1)

where $x = r/r_c$ and $y = (v(r)/v_c)^2$. Our method of solving eq. (1) is as follows. Starting at x=1, the solution is known (by inspection) to be y=1. Then, we at first move outward in radius from x=1, and select a series of increasing values of *r* (i.e. values of x) extending outwards from $r_c$ (i.e. x=1) to $x=200r_*/r_c$. At each value of x, we solve eq. (1) for y, and then, knowing the numerical value of $v_c$ in the corona which we are dealing with (i.e. a corona with a specified *T*), we obtain the solution for *v(r)* = $v_c\sqrt{y}$ in units of km/sec at that particular radial location. After completing the outward solution, we return to x=1, and then select a series of decreasing values of *r* extending inwards from $r_c$ towards the surface of the star. For each x, we again solve eq. (1) to obtain *v(r)* at each radial location. As we approach the star, y approaches zero, and the logarithmic term on the left-hand side of eq. (1) diverges. As a result, there are numerical difficulties in finding a solution near the surface of the star. In practice, we find that we cannot carry the solution inward of $r \approx 1.1$-$1.2r_*$. The solutions we have found for *v(r)* in coronas with *T* = 1 and 2 MK replicate well the curves plotted by Parker (1963: his Figure 6.1), and we do not plot them here. Once we know *v(r)* for a corona with temperature *T*, we calculate the local values of density $n(r) = C/r^2 v(r)$ at each radial location.

3.2. The magnetic field profile, including the "source surface"

The magnetic field around a flare star consists in principle of two components. (i) Active regions contribute strong fields which, because they are high order multipoles, are confined to short distances close to the star. (ii) The global field, which is the lowest order multipole, i.e. a dipole, provides the dominant contribution to the total field at large distances. Both (i) and (ii) have been included in the works of Mann et al (2003) and of Warmuth and Mann (2005) by representing an active region as a dipole which is centered underneath the stellar surface and oriented either horizontally or vertically relative to the surface. In these works, the dominant effects of active region fields are readily perceived within (roughly) 0.5 stellar radius of the surface. **At radial distances of 1.5R(star) and more, Mann et al and Warmuth and Mann find that the global field dominates. In the calculations of $V_a$ to be presented in the main part of this paper, we include only component (ii). Clearly, our approach represents an oversimplification of the field strength at any particular radial distance from a flare star. However, towards the end of the paper (in Section 8), we will consider briefly a more complete scenario,**

**including components (i) and (ii), and we shall attempt to estimate the distance at which the global field strength will dominate over the active region field.**

The large-scale field of the Sun is the ultimate origin of the fields which can be measured by *in situ* detectors carried by spacecraft in the solar wind. If we imagine moving from the surface of the Sun outward into interplanetary space, the field will undergo a transition from predominantly closed to predominantly open at a certain radial distance from the Sun. The place where the transition to almost entirely open field lines occurs is referred to as the "source surface" (SS: see Schatten et al 1969). In calculating the field in the solar wind, one assumes that outside the SS, the interplanetary field is radial. Depending on the radial location *r(SS)* where the SS is located, one can obtain a more or less good correlation between the fields on the solar surface and the field strengths in interplanetary space. Koskela et al (2017) have reported that, in the case of a 40-year data set from the Wilcox Solar Observatory, they obtain the best matching between the polarity of the coronal field and the polarity of the field measured at 1 AU if they choose *r(SS)* = 3.25 $R$ : with that choice, the polarity match reaches its highest value, about 79%. However, the maximum in the polarity match is rather insensitive to the choice of *r(SS)*: the polarity match is found to decrease by only 2% (from 79% to 77%) if *r(SS)* is moved from 3.25 out as far as 5.75 $R$ . These results suggest that it may be difficult to narrow down significantly the empirical value of *r(SS)* to better than a few solar radii.

In this paper, we will assume that, when we wish to calculate $V_a$ at a radial location which lies inside the SS, the total field strength *B(r)* will scale as $1/r^3$, as befits a dipole. But at radii *r > r(SS)*, we will use the scaling *B(r) ~ $1/r^2$*, as befits a radial field. We shall refer to this field distribution as a "hybrid" magnetic model of the corona/wind.

We note that Mann et al (1999), in their discussion of the field from the "quiet Sun" (corresponding to what we refer to as the global dipole) assume that the magnitude of the field *B(r)* scales as *$1/r^2$* at all radial locations, including the inner corona, all the way in to the solar surface. In the present paper, for the purposes of checking our results against those of Mann et al (2003), we will initially adopt *B(r) ~ $1/r^2$*. But in later parts of the paper, we will incorporate the profile *B(r) ~ $1/r^2$* for the field magnitude only at radial locations *r* which lie outside *r(SS)*.

In order to proceed, we need to choose a value for *r(SS)*. To do this, we assume that, as we proceed outward from the stellar surface, the field becomes radial (i.e. "opens up") at the radial location *r* where the ram pressure of the wind $P_w(r) = 0.5\, n(r)\, m_p\, v(r)^2$ becomes as large as the local magnetic pressure $P_m$ = $B(r)^2/8\pi$. We note that a different criterion for determining the value of *r(SS)* was originally suggested by Schatten et al (1969), namely, the location where the pressure of the transverse magnetic field component equals the thermal gas pressure. The latter criterion leads to *r(SS)* values which lie rather close to the Sun, no more than 0.7$R$ above the photosphere. The latter value seems to be too small to be consistent with the results reported by Koskela et al (2017). We shall find below that when we apply our criterion $P_w(r) = P_m(r)$, the numerical values of *r(SS)* will be found to be more consistent with those reported by Koskela et al (2017).

4.  **ALFVEN SPEED PROFILE IN A PURELY RADIAL MAGNETIC FIELD**

In order to set the stage for calculating the Alfven speed in flare star coronae, we first refer to the case of the Sun. Mann et al (2003) have plotted values of $V_A(r)$ in the case where the isothermal solar corona has a temperature $T = 1.0$ MK. The solar wind has a critical point at a radius $r_c = GM/2v_c^2 = 6.91\,R$ for $T = 1.0$ MK. Solving eq. (1) above, we confirm that at $r = 215\,R$ (i.e. 1 AU), $v(r)$ has a value of 427 km/sec, as listed by Mann et al (1999) for $T = 1$ MK. Using $C = 6.3 \times 10^{34}$ per second, the profile of $v(r)$ then leads to the required profile of $n(r)$.

In order to complete the calculation of $V_A(r)$, we follow the choice of Mann et al (2003) where the quiet Sun field strength is taken to be $B_{qs}(r) = B_S(R/r)^2$, i.e. the field is radial all the way from the solar surface out to radius $r$. Mann et al (2003) adopt a surface field $B_S = 2.2$ G. This process leads to the profile for $V_A(r)$ plotted as the lowest curve in Figure 1.

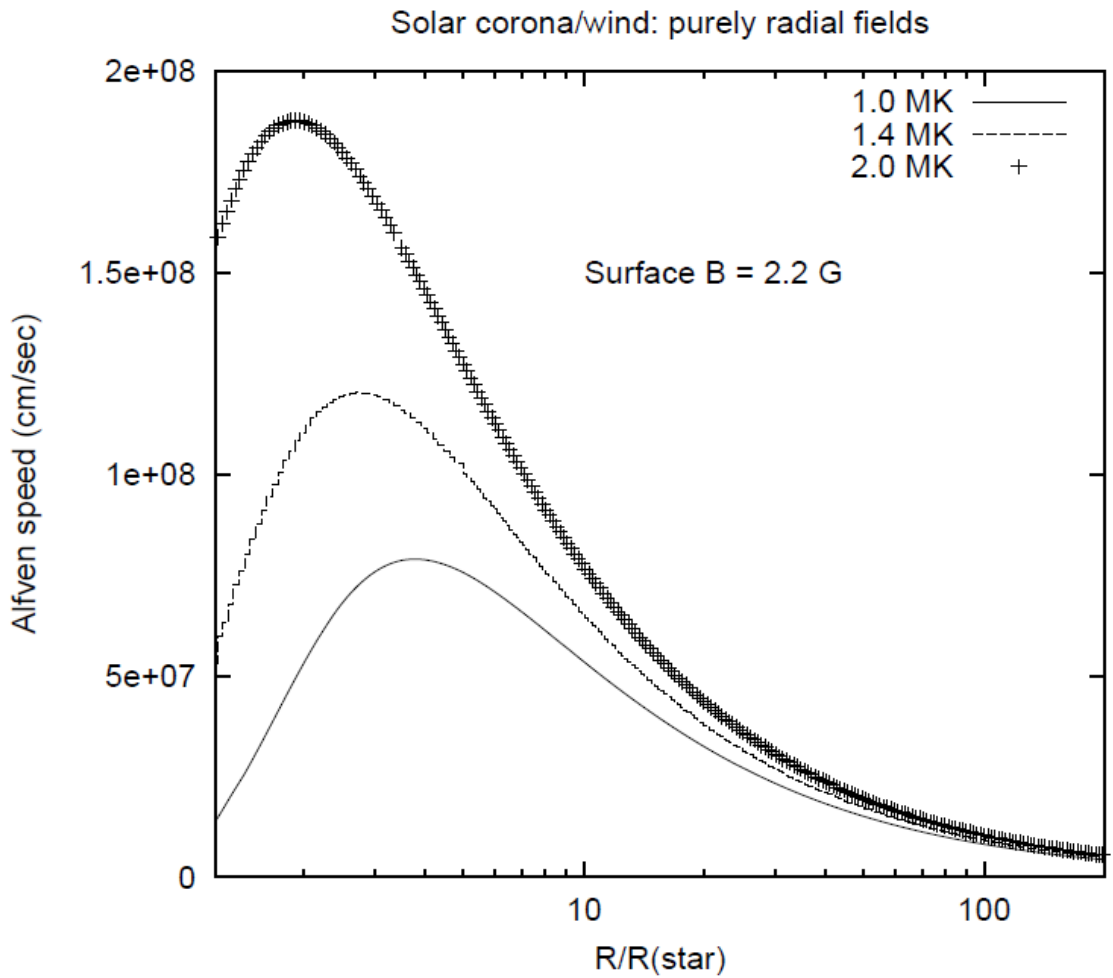

Figure 1. Radial profiles of the Alfven speed in the solar corona/wind for three different values of the coronal temperature. In this figure, the field on the solar surface has a strength of 2.2 G, and the field is assumed to be purely radial (i.e. open, monopolar) at all radii. The continuous line in this figure (i.e. $T = 1$ MK) agrees with the $V_A(r)$ profile published by Mann et al (2003).

Comparing with the results of Mann et al (2003: the dotted curve in their Figure 6), we note the following quantitative agreements between our results and those of Mann et al (2003): $V_A(r)$ has a small value close to the solar surface, rises to a peak value of 700-800 km/sec at a radial distance of about $4R$, and then declines at larger distances, falling to 50-100 km/sec at about 0.5 AU.

Also in Fig. 1, we illustrate the profiles of $V_A(r)$ for two other coronal temperatures: 1.4 MK and 2.0 MK. The first of these (1.4 MK) is the temperature which fits best the density scale height of the Newkirk (1961) model. The choices of upper and lower limits on the temperature of the quiet corona (2.0 MK, 1.0 MK) are plausible in the presence of the physical processes which are associated with the combined operation of radiative and conductive energy losses (Mullan 2009). The temperatures of 1.4 and 2.0 MK are reported by Mann et al (1999) as having critical radii of 4.90 and 3.43 $R$. We confirm that our results for these models (not shown in Fig. 1) yield wind speeds at 1 AU of 528 and 674 km/sec respectively, in agreement with the results listed by Mann et al (1999).

**An important feature of the results in Fig. 1 is that the values of $V_A$ become systematically *larger* as we select larger values for the coronal temperature. The reason for this behavior has to do with the assumptions we adopted in Section 3.1: in order that the model we use for the corona satisfies the observed mass loss rate from the Sun (2 x $10^{-14}$ $M$ /yr), an increase in coronal temperature (with its faster wind speed) must be inevitably accompanied by a reduction in the overall density profile. This behavior can be seen explicitly in Table 1 of Mann et al (1999): when they assigned coronal temperatures of 1.0, 1.4, and 2.0 MK, they had to adopt the following values for the particle number density at the base of the corona: 5.14 x $10^9$, 1.61 x $10^8$, and 1.40 x $10^7$ cm$^{-3}$ respectively in order to agree with the observed mass loss rate. Clearly, for a given field strength, the reduced density at the base of the corona when $T$ is hotter leads to larger values of $V_A$ in the hotter models (as seen in Fig. 1). We will return to this aspect of our results when we consider the coronae in flare stars (see Section 6 below).**

Of course, if the actual surface field $B_S$ in the Sun is not equal to 2.2 G, the numerical values of $V_A(r)$ in Fig. 1 should be multiplied by factors of $B_S/2.2$.

## 5. ALFVEN SPEED PROFILE IN A HYBRID MAGNETIC MODEL: THE SOLAR CASE

Now we consider a model of the solar corona/wind which includes a source surface. For a given coronal temperature, we use the same velocity and density profiles as we used in Section 4. We also use the same value of $B_S$ = 2.2 G as Mann et al (2003) used for the surface field strength. But now we use a $1/r^3$ profile for the field for $r<r(SS)$, and switch to a $1/r^2$ profile for $r>r(SS)$. The location of the SS is determined from the condition $P_w(r) = P_m(r)$.

The Alfven speed profiles which we obtain for three coronal temperatures are shown in Figure 2.

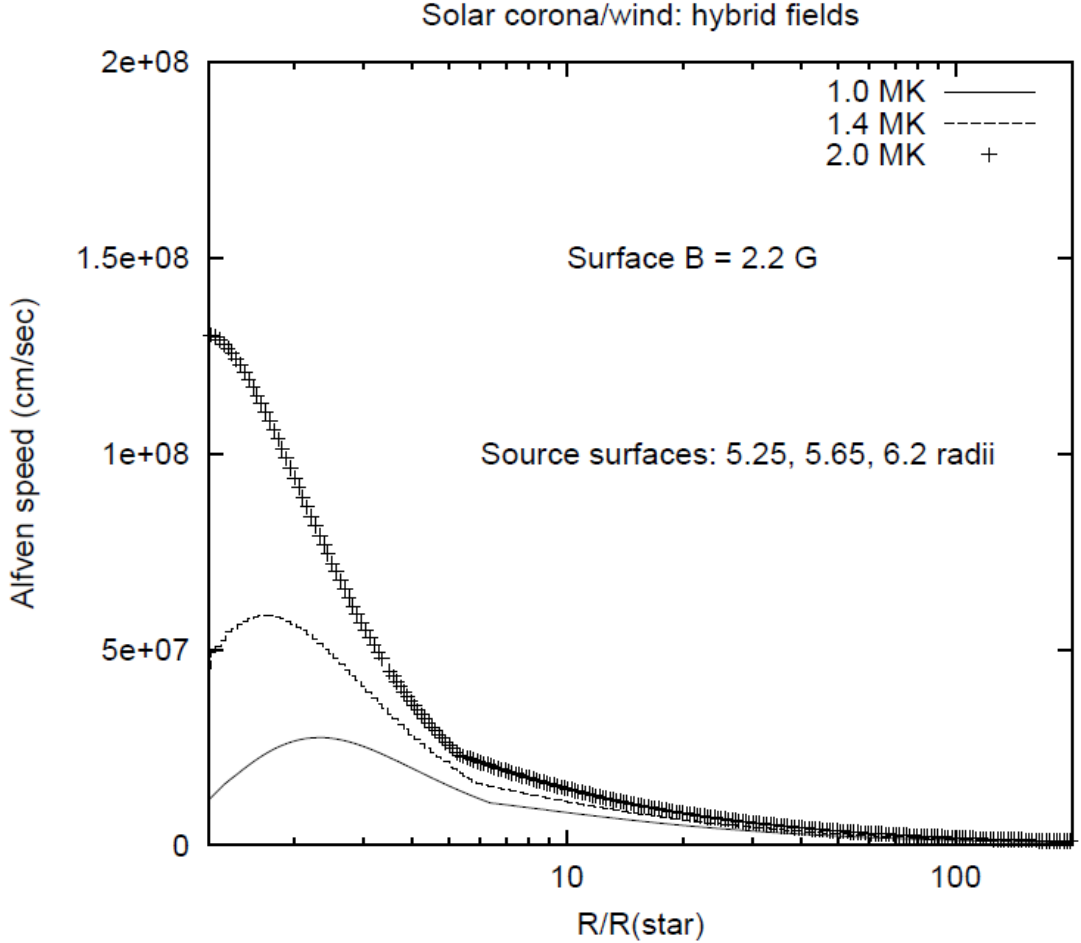

Figure 2. Radial profiles of the Alfven speed in the solar corona/wind for three different values of the coronal temperature. In this figure, the field on the solar surface has a strength of 2.2 G, and the field has a hybrid structure: the field is assumed to be dipolar inside the source surface, and monopolar outside the source surface. The radial locations of the source surfaces are listed in the Figure.

Qualitatively, the $V_A$ (r) profiles in Fig. 2 resemble those in Fig. 1, i.e. $V_A$ is smallest in regions close to the Sun, as well as in regions far from the Sun, and there is a peak in $V_A$ at intermediate radial locations. But the $V_A$ (r) curves in Fig. 2 are not as smooth as those in Fig. 1: on each curve in Fig. 2, one can spot a change in slope at the radial location of the source surface, r(SS). We find that r(SS) occurs at 6.2R for the coolest corona (T = 1 MK), at 5.25R for the hottest corona (T = 2 MK), and at 5.65R for the intermediate case (T = 1.4 MK). In the cases T = 1.4 and 2 MK, the values we have found for r(SS) overlap with the range of empirical values of r(SS) which have been reported by Koskela et al (2017) as providing the best polarity fits between the Sun's fields and the interplanetary fields over a span of almost four solar cycles. We consider that the overlap of our *r(SS)* values with the empirical results points to a fair degree of physical plausibility in our hybrid model.

An important difference between the $V_A$ values in Fig. 2 and those in Fig. 1 is that the speeds in the hybrid model are systematically smaller than those in the purely radial model. The reason is readily apparent: in the hybrid model, the field in the inner corona falls off with increasing *r* more rapidly than the fall-off which occurs in the purely radial model. As a result, even though the surface fields are identical in Figs. 1 and 2, the field strength at any given *r* outside the Sun is systematically weaker in Fig. 2 than that in Fig. 1.

The results in Figure 2 demonstrate that an MHD disturbance which propagates upwards through the corona will excite a Type II radio burst in a 1 MK corona if its outward speed exceeds roughly 250 km/sec. And in regions of the corona where the temperature is as high as 2 MK, excitation of a Type II burst requires a disturbance to move at speeds of at least 1300 km/sec.

Are these critical speeds of any relevance in the context of solar data? Yes: as an illustration, we note that in the general population of CME's (which are typically associated with Type II burst excitation), the average speed is ~480 km/sec (Gopalswamy 2006). In a separate study, Gopalswamy et al (2009) report that for CME's which occur in the LASCO data, the average speed of CME's is 779 km/sec, while CME's associated with Type II bursts have average speeds of 534, 808, and 1367 km/sec depending on the wavelength range of the burst.

We may also cite a survey of 22 CME's which were observed between 2008 and 2012. Mostl et al (2014) have reported on initial and propagation speeds which range from as slow as 260 km/sec to as fast as 2715 km/sec. Although Mostl et al do not explicitly refer to Type II radio data, our results in Fig. 2 predict that the fastest CME's reported by Mostl et al should certainly be able to excite Type II radio bursts even if the CME moves through a region of the corona where the temperature is as hot as 2 MK. And even their slowest CME's might be able to excite Type II's if they were to propagate through a region of the corona where the temperature is as low as 1 MK. In principle, the results plotted in Figure 1 suggest that Type II bursts could be readily excited by the fastest CME's in the list of Mostl et al (2014), but the slowest CME's in their list should not be accompanied by Type II bursts. However, we consider the results in Figure 2 to be more physically meaningful than those in Fig. 1.

Once again, if the actual $B_S$ in the Sun is different from 2.2 G, then all values of $V_A(r)$ in Fig. 2 should be multiplied by factors of $B_S/2.2$. If, e.g. $B_S$ were to be in fact as large as (say) 6.6 G, CME's propagating through coronal regions where *T* = 1 (or 2) MK will generate Type II radio bursts only if the CME speed exceeds 750 (or 4000) km/sec. Such changes would lead to the absence of Type II emission from at least some, and possibly all, of the CME's in the lists of Gopalswamy et al (2009) and of Mostl et al (2014).

## 6. ALFVEN SPEED PROFILE IN A HYBRID MAGNETIC MODEL: THE FLARE STAR CASE

Turning now to flare stars, in order to calculate the Alfven speed profile, we need to choose appropriate values of mass loss, coronal temperature, and field strength. We have already indicated (Section 3.1) that if we use the same mass loss constant which is valid for the Sun, we may expect that our coronal/wind densities in flare stars should be within a factor of a few of the solar values.

**As regards the coronal temperature *T* in flare stars, we need to refer to spacecraft data. The earliest estimates of *T* in a sample of late-type stars were provided by the EINSTEIN X-ray detectors: with a very limited number of energy channels, the resulting temperatures were subject to considerable uncertainty. Nevertheless, in a study of 45 G, K, and M stars, Katsova et al (1987) analyzed the EINSTEIN data and found 5 stars have *T* = 1 + *f* MK (where *f* <1), 17 stars have *T* = 2 + *f* MK (where *f*<1), and 23 stars have *T* > 3 MK. That is, about 50% of the stars had *T* < 3 MK. More recent X-ray data, consisting of high resolution spectra obtained by XMM-Newton and by Chandra, have allowed the derivation of emission measure distributions (EMDs): by weighting these EMDs for a sample of 24 low-mass main sequence stars, a more physically reliable average coronal temperature has been derived for each star (Johnstone and Gudel 2015). The lowest *T* value, referring to the Sun at solar minimum, is 0.97 MK. Two stars have *T* = 1-2 MK. Six stars have *T* = 2-3 MK. The remaining targets have *T* > 3 MK, with the hottest corona (47 Cas B) having *T* = 10.72 MK. Thus, in this case, the fraction of stars with *T* < 3 MK was only about one-third. Apparently the EINSTEIN detectors were not especially effective at identifying the hotter coronae.**

**In view of these results, the question arises: what values should be adopt for the coronal temperatures in flare stars? At this point, we recall an important result which emerged from our results in Figure 1 (see Section 4, penultimate paragraph): higher *T* values result in larger values of $V_A$ . If, therefore, we were to adopt *T* = 1-2 MK for our calculations of flare star winds, then we would obtain firm *lower limits* on the $V_A$ which would be appropriate if we were to insert the actual empirical values of *T* (i.e. as large as 3-10 MK). Thus, whatever conclusions we will draw as regards the unlikelihood of CME speeds exceeding the $V_A$ values in flare star coronae (see Figure 3 below) would actually be reinforced if we were to insert the (higher) empirical values of *T*. And as we shall find, the $V_A$ values which we shall derive using *T* = 1-2 MK (see Figure 3) are already so close to the speed of light that there is physically very little room for increasing $V_A$ anyway.**

**As a result, and in order to facilitate a comparison between our solar results and flare stars, our goals in this paper will not be materially affected if we retain the values *T* = 1.0, 1.4, and 2.0 MK in our study of flare stars.**

As regards magnetic field strengths, it has been known for some time that as one observes at later spectral types on the main sequence, the average field strengths on the surface are found to increase (e.g. Saar 1996). This leads us to expect that the fields which are pertinent in the coronae/winds of flare stars are in all likelihood stronger than those in the solar corona/wind. How much stronger are the flare star fields likely to be than in the Sun? The field strengths on certain flare stars have recently been reported to be as strong as 7 kilogauss (Shulyak et al 2017) and at least 5.1 kG (Berdyugina et al 2017): the latter feature was reported to occupy at least 11% of the visible hemisphere. The features in which these strong fields occur may be cool spots: if so, and if these are analogous to sunspots, the reported fields are probably caused by toroidal flux tubes breaking through the surface in localized structures. Such toroidal fields give rise to multipolar components of the stellar field, and these fall off rapidly with increasing distance from the star (Mann et al 2003; Warmuth and Mann 2005). These fields are unlikely to contribute significantly to the fields in the corona and wind of a flare star.

Instead, in order to calculate realistic Alfven speeds in the corona/wind, we need to have information about the *poloidal* fields which create the largest-scale fields on the star, i.e. those which give rise to a global dipole. How can the strengths of such fields be estimated?

In a series of papers dealing with quantitative modeling of the inhibition of the onset of convection in the presence of a vertical magnetic field (Mullan and MacDonald 2001; MacDonald et al 2018, and references therein), it has been found that the global parameters (the radius, the effective temperature, the age) of various low-mass stars can be replicated provided that vertical magnetic fields $B_v$ of certain strengths are present on the surface of the star. In the cases of the flare stars GJ65A and B (with spectral types dM5.5 and dM6), these magneto-convection models predict that $B_v$ = 1800 and 2250 G respectively (MacDonald et al 2018). And in the case of the flare star Trappist-1, the models predict that $B_v$ is in the range 1450-1700 G (Mullan et al 2018).

High resolution images of the solar corona during total eclipses which occur near the times of solar minimum (e.g. Habbal et al 2010, especially their Fig. 1) provide evidence that the vertical component of the field is what dominates in the polar regions of the Sun. In such regions, toroidal fields probably contribute little to the local magnetic field. In view of this, we consider it physically plausible to consider the following hypothesis: the value of $B_v$ can serve as a proxy for the surface field strength of the global dipole of a star. As a test of this, we note that in the case of the Sun, it allows us to predict the changes in *p*-mode frequencies, as well as the change in solar radius, associated with the solar cycle: these predictions are found to be consistent with the available data (Mullan et al 2007).

On this basis, the principal suggestion in the present paper is the following: in a flare star, an appropriate choice for $B_S$ would be the value of $B_v$ in the magneto-convective model which best fits the global parameters of that star. Given the above values of $B_v$ in GJ65A, GJ65B, and Trappist-1, we consider that $B_S$ values of order 1 kG would not be outside the realm of possibility for a flare star.

With this choice, we show, in Figure 3, the profiles which we obtain for $V_A(r)$ using three values of coronal temperature.

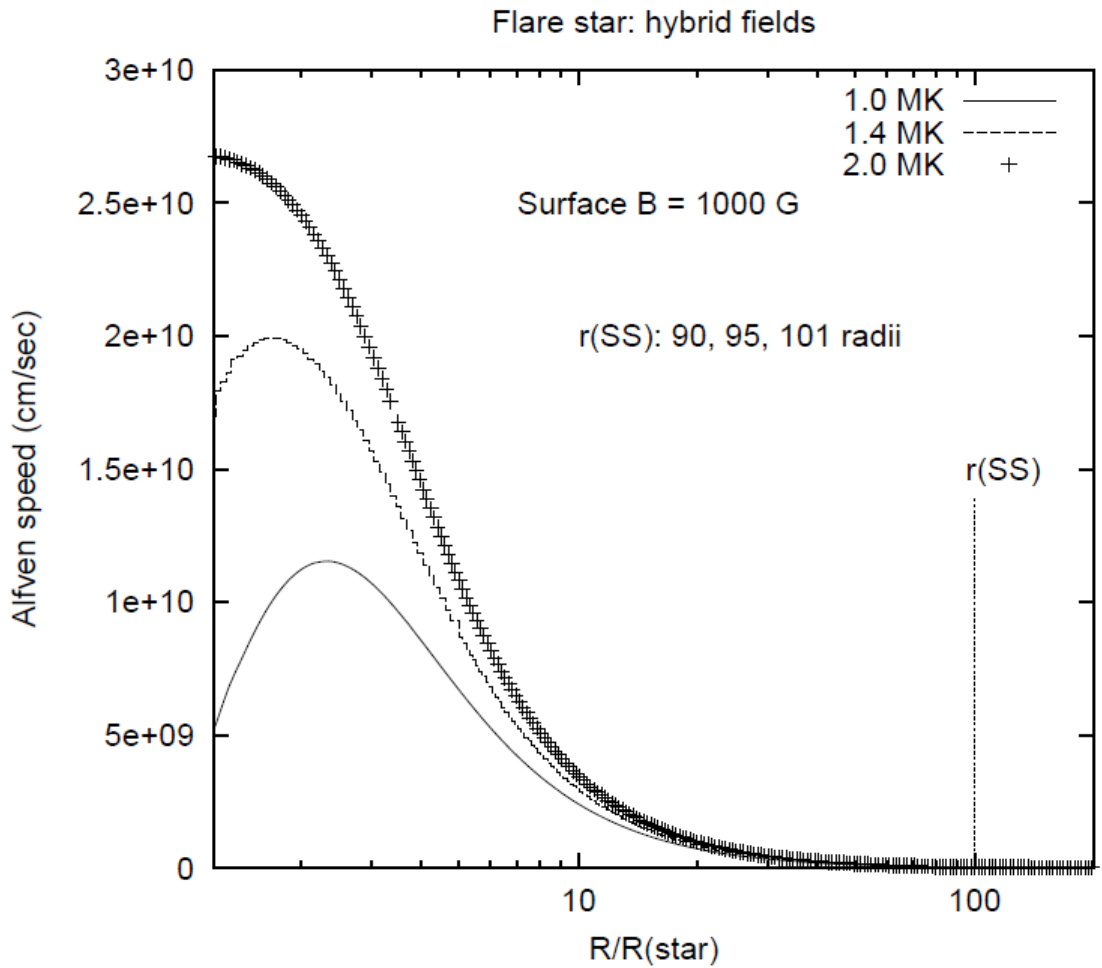

Figure 3. Radial profiles of the Alfven speed in the corona/wind of a flare star for three different values of the coronal temperature. In this figure, the field on the stellar surface has a strength of 1 kiloGauss, and the field has a hybrid structure: the field is assumed to be dipolar inside the source surface, and monopolar outside the source surface. The vertical line labelled *r(SS)* indicates roughly the locations of the source surfaces.

There are two significant features to be noted about Fig. 3. First, the field is so strong that the source surfaces are located much farther from the star than in the cases plotted in Fig. 2: for the flare star, we find that the *r(SS)* values lie at distances of 90-100 stellar radii, where it is not easy for the reader to see the change in slope which accompanies passage through *r(SS)*. Second, the numerical values of $V_A(r)$ are close to the speed of light. (In fact, if we were to mistakenly use the non-relativistic formula for $V_A$, we would find Alfven speeds which would be unphysical, i.e. clearly in excess of c.) In stars for which Figure 3 provides a realistic assessment of $V_A(r)$, we would have to conclude that CME's would essentially never have speeds which would be definitely in excess of $V_A$. Therefore, in such stars, Type II radio bursts could not be excited. CME's might very well exist, but they would be radio-quiet.

An anonymous referee has raised an interesting possibility for Type II emission from the distant outer wind of a flare. Although we have found (see Fig. 3) that the largest values of $V_A$ in a flare star wind may be close to the speed of light *in the inner wind,* this is no longer the case at great distances. Specifically, Figure 3 illustrates that at radial distances of 20-30$R_*$, the $V_A$ values have decreased to less than 10,000 km/sec. Suppose, therefore, that a CME were ejected from a stellar flare with a speed of 10,000 km/sec: this clearly exceeds the fastest CME (almost 3000 km/sec: Yashiro et al 2004) observed in the Sun, but larger energies in stellar flares may help to propel CME's from stellar flares at larger speeds than those from solar flares. If the CME from a stellar flare could maintain a speed of 10,000 km/sec out to radial distances of 20-30$R_*$, an MHD shock might be generated out there and could lead to Type II emission. Observations of solar bursts (Al-Hamadani et al 2017) certainly confirm that Type II emission occurs in the solar wind at radial locations which are as far out as 20-30$R_*$ . The relevant frequencies of solar Type II emission at such radial distances are observed to be in the range 0.2-2.2 MHz, and are too low to be detected from the ground.  In the case of a flare star, the corresponding plasma frequencies (f[MHz] ≈ 0.01 √n(r) where n(r) is in units of $cm^{-3}$) at equivalent radial distances might in some cases be large enough to be detectable by ground-based telescopes. However, with ionospheric cutoff frequencies in the range 5-10 MHz, ground-based observations will be possible only if the Type II emission originates in a medium where n(r) > (2.5-10)x $10^5$ $cm^{-3}$ . Inspection of our flare star model in which *T* = 1 MK and the mass loss rate is equal to the solar value, indicates that at r = 20-30$R_*$ , the number densities are in the range 530-1320 $cm^{-3}$. These lead to f(MHz) < 0.4 MHz, requiring space-borne detectors. If Type II bursts from the outer winds of flare stars are to be detectable from the ground, the wind densities will have to be larger than the values we adopt in this paper: this suggests that flare stars with the largest mass loss rates, such as ε Eri (Wood 2018), might be favorable targets.

## 7. ALFVEN SPEED PROFILES FOR STARS WITH INTERMEDIATE MAGNETIC FIELD STRENGTHS

The flare stars which we referred to in Section 6 in connection with global field strengths have masses of order 0.1 *M* : these masses are among the smallest values for which magneto-convective models have been computed. In this regard, their global field strengths may not be altogether typical of other low-mass stars. For example, in a survey of 15 low-mass stars with known ages for which magneto-convective models have now been computed, MacDonald and Mullan (2017) have reported that among these stars, with masses ranging from 0.1 to 0.6 *M*  , the mean surface field value (taken as the arithmetic mean of the values listed for each star in Table 1 of that paper) has a smallest value of 290 G.

In order to give representative results for stars where the magnetic fields are stronger than the solar value, but not as strong as those in the prominent flare stars cited in Section 6, we have therefore calculated $V_A(r)$ profiles for a star where the surface field is chosen to have the value $B_S$ = 290 G. Once again, we have calculated three versions of the $V_A(r)$ curves, with coronal temperatures as before equal to *T* = 1.0, 1.4, and 2.0 MK. Unsurprisingly, the profiles (not shown) turn out to be intermediate between the results in Fig. 2 and those in Fig. 3.

Thus, for the $T = 1.0$ MK model, we find that the source surface lies at $56R_*$, and $V_A$ has a maximum value of 36,000 km/sec at a radial distance of $2.3R_*$. For the $T = 2.0$ MK model, the source surface is found to lie at $49R_*$, and $V_A$ has a maximum value of 150,000 km/sec at a radial distance of $1.2R_*$.

These maximum values of $V_A$, although smaller than those shown in Fig. 3, are nevertheless so large that CME's from a star with $B_s$ no greater than 290 G must be ejected with speeds in excess of several tens of thousands of km/sec if Type II radio bursts are to be excited. Of course, it is known that speeds of up to 30,000 km/sec can be associated with supernova events when an entire star explodes. But in our case, where we are considering events in which material is ejected from a localized magnetic portion of a stellar surface, such events can hardly be expected to replicate the speeds of ejecta from supernovae.

We conclude that Type II radio burst excitation is difficult to achieve in the inner corona/wind even in stars where the surface field is no more than 2-3 hundred Gauss.

## 8. WHERE DO GLOBAL FIELDS DOMINATE ACTIVE REGIONS FIELDS?

So far, we have confined our attention to only the global field. However, to be realistic, the Sun and flare stars contain not only a global field, but also contain active regions (AR), i.e. localized regions of strong fields on their surface. How will our results for $V_A$ be affected by the presence of AR fields? We are especially interested in evaluating the radial distance beyond which the global field dominates. We address this issue in a simplified manner as follows.

To address the magnetic effects of an AR, it is customary to represent the AR by a magnetic dipole with a certain polar field strength submerged at a finite depth $D$ below the surface (Mullan 1979; Mann et al 2003; Warmuth and Mann 2005). The orientation of the dipole relative to the star's surface can be either radial (giving rise to a single spot on the surface) or azimuthal (giving rise to a bipolar pair of spots on the surface). For the solar case, values of $D$ in the literature range from (0.05-0.3)$R_*$ (Mullan 1979) to $0.1R_*$ (Mann et al. 2003), to only 10-35 Mm (i.e. 0.014-0.05$R_*$) (Warmuth and Mann 2005). The reason for such a wide range of $D$ values has to do with the amount of information which was available concerning the depth of spots. In 1979, helioseismology had not progressed to the state where depths of sunspots could be determined reliably. As a result, Mullan (1979) adopted values of $D$ which allowed the spots to lie at depths anywhere within the solar convection zone. As regards the strength of the dipole, Mullan (1979), who was interested in flare stars, selected a surface field strength of $B_s = 10^4$ G at the center of the spot. But Warmuth and Mann (2005) citing results from solar work, chose $D$ as small as 5-6 Mm or 9-12 Mm, depending on the data set. And Warmuth and Mann assigned surface field strengths of $B_s = 2500\text{-}2900$ G to the AR's. Once values are chosen for $B_s$ and for $D$, an upper limit on the field strength at any height H above the surface can be calculated: $B_{AR}(H) = B_s [D/(D+H)]^3$. This will be the field strength at a height $H$ on the extended radius vector joining the center pf the star and the dipole. At the same position, the global dipole will have a strength of $B_p(H) = B_{ps}(R/R+H)^3$ where $B_{ps}$ is the dipole field strength on the surface of the star at its pole.

Adopting the solar field strengths listed by Warmuth and Mann (2005), i.e. $B_s = 2700$ G for the AR and $B_{ps} = 3.4$ G for the global dipole, we find that the AR field strength is expected to be comparable to the global field strength at a radial location $R = R_c$ where $(1 + H/R)^3 = (B_{ps} / B_s)*(1+H/D)^3 = 0.0013(1+H/D)^3$. That is, $1+H/D = 9(1+H/R)$. Setting $D = xR$, where $x<1$, and defining $y = H/R$, we can solve for $y$ once a value is chosen for $x$. In the lower limit on $D$ considered by Warmuth and Mann (2005), i.e. $x = 0.014$, the solution is $y = 0.13$. In this case, the AR field becomes weaker than the global field at heights above $0.13R_*$, i.e. outside a radial location of order $R_c = 1.13R_*$. If we use the upper limit on $D$ considered by Warmuth and Mann (2005), i.e. $x = 0.05$, the solution is $y ≈ 0.7$. In this case, the AR field becomes weaker than the global field at heights above $0.7R_*$, i.e. outside a radial location of order $R_c = 1.7R_*$.

A check on these conclusions can be obtained by inspecting Figures 6-8 of Warmuth and Mann (2005). It can be seen that the largest magnetosonic speeds are indeed confined to a localized region very close to the AR: $r < 1.1R_*$. However, it appears that the field transitions over into what is essentially a global dipole structure at greater radial distances: within the range of radial locations plotted by Warmuth and Mann (i.e. from 1.0 to 2.0 $R_*$), visual inspection of their Fig. 6-8, which use a value of $D$ that is intermediate between the upper and lower limits considered in the previous paragraph, suggests that the transition occurs no farther out than roughly $1.5R_*$, i.e. within the range $1.1-1.7R_*$ estimated above. This comparison suggests that the simplified approach described above may provide fairly realistic estimates of the radial location where the global dipole of the Sun begins to dominate over the AR fields.

What about the non-solar case, e.g. flare stars? In order to estimate the value of $R_c$ the transition from AR field to global field occurs in such stars, we need to assign appropriate numerical values to the two key field strengths $B_s$ and $B_{ps}$. Empirical data indicate that fields as strong as 7 kG exist on the surface of flare stars (Shulyak et al 2017). Therefore, a choice $B_s ≈ 10^4$ G, as chosen by Mullan (1979), does not seem inappropriate. What about the global dipole field strength in flare stars? In a study of mid-M flaring dwarfs, Morin et al (2008) found that the fields consist of an axisymmetric global dipole which contains a fraction $f = 56-79\%$ of the field energy. With a mean surface field strength of $<B> = 0.18-0.78$ kG, this means that the global dipole is estimated to have a strength on the stellar surface of $B_{ps} = (0.75-0.89)<B>$, i.e. a strength of up to 0.7 kG. Clearly, these global fields on flare stars are much larger than in the case of the Sun. Is there any confirmation for such strong global fields? Independent modeling of the global structure of active stars, using a magneto-convective theory based on a quantitative criterion for the inhibition of convective onset in the presence of a magnetic field (MacDonald et al 2018; Mullan et al 2018) suggests that the global field strengths on the flare stars Trappist-1 and GJ 65A/B are of order $B_{ps} = 1$ kG on the surface of the star at the poles. Thus, even though the numerical values are certainly not as reliably known as in the solar case, we might be able to derive a crude estimate of $R_c$. Proceeding as above, we now have $(1 + H/R)^3 = (B_{ps} / B_s)*(1+H/D)^3 = 0.1(1+H/D)^3$. This leads to $1+H/D = 2(1+H/R)$. What value should we use for $D$? The convection zones are certainly deeper in flare stars than in the Sun: so the values of $x = D/R$ may be larger than in the case of the Sun. However, we cannot place the dipole as deep as the center of the star, where x would have to be set at x=1: in such a case, inspection of the last equation above shows that no solution is possible. So how deep should we place the equivalent dipole for an AR on a flare star? In this regard, we note that Mullan (1979) found that the dimming of the star's light due to a spot rises to 0.34 magnitudes if the spot is modeled by a horizontal dipole placed at a depth $D = 0.3R_*$: such a dimming is not far from the largest observed amplitudes of rotational modulation in the active system BY Dra. Therefore, setting x = 0.3, we have that $1 + 3.3y = 2(1+y)$, i.e. $y = 0.7-0.8$. That is, the transition from

**AR field to global field occurs at a radial location $R_c$ of order 1.7-1.8 $R_*$ . Interestingly, although the magnetic parameters of flare stars and Sun are quite different, our solution for $R_c$ is not greatly different from the solution obtained in the solar case when we use the upper limit on *D* suggested by Warmuth and Mann (2005).**

**In summary, based on the simplified approach outlined here, we expect that the global field will dominate the AR fields in the corona/wind of a flare star once we are outside a critical radial location which is no farther out than *r* = 1.7-1.8$R_*$. Returning to Figure 3 above, in which we assumed that only the global field is present in the corona/wind, we can see that this assumption is valid for almost the entire range of radial locations plotted along the horizontal axis. Specifically, the assumption of global field dominance breaks down only within the left-most 8% of the horizontal axis. Thus, our assumption is probably acceptable over 92% of the range of (log) radii plotted in Fig. 3.**

## 9. CONCLUSION

In the solar corona, where the global surface fields have strengths of a few Gauss, we find that the observed speeds of CME ejecta (~300 – 3000 km/sec) can in many cases exceed the coronal Alfven speed $V_A(r)$. As a consequence, it is not surprising that Type II radio bursts (which require Alfvenic Mach numbers of 1.1-1.2 or more to be emitted) can be excited by CME's quite often in the solar corona/wind, especially around times of solar maximum. Because of the magnetic properties of the Sun, with its rather weak global field (no more than several G), it can readily be understood why CME's ejected in association with solar flares are quite effective at generating Type II radio bursts.

But in flare stars, the situation is different. In this paper, we compare and contrast the physical conditions in the corona/wind of flare stars relative to the corona/wind of the Sun. Our work is based on recent magneto-convective models of flare stars which suggest that vertical magnetic fields on the surface of certain flare stars may have strengths of order 1 kilogauss. Assuming that the global field of a flare star is associated with vertical fields of such magnitude, and using density profiles which are consistent with the known mass loss rates of K and M dwarfs, we find that $V_A$ in the corona/wind of such a star approaches the speed of light. In such a situation, Type II radio bursts, which require a disturbance (such as a CME) to move through the corona at a speed in excess of $V_A$ in order to be excited, cannot be generated.

Even In certain low-mass stars where the global field may have surface strengths of "only" 200-300 G, we find that Type II radio bursts are not easily excited. In such stars, Type II bursts will be excited only if the CME's propagate at speeds faster than roughly 10% of the speed of light.

To be sure, the larger energy of certain stellar flares compared to solar flares may very well lead to larger kinetic energies of CME's in flare stars than in the Sun. But even in stellar flares where the total energy released in the flare exceeds the largest solar flare energies by factors of as much as 100-1000, and where the expected CME speeds may be larger than in the Sun by factors of 10-30, it will be a rare event in which the CME speed will approach the maximum values we have found for the Alfven speed. Even less likely, therefore, is it that the CME speed will actually exceed the maximum coronal values of $V_A$, thereby satisfying the criterion for excitation of a Type II radio burst.

This leads us to the conclusion that, even in very large stellar flares, except in the most extreme conditions, the CME's which are created probably cannot fulfil the criterion required for Type II burst excitation. We suggest that such events may correspond to the "radio quiet" CME's mentioned by Villadsen and Hallinan (2018).

The magnetic conditions in flare star coronae are so different from what occurs in the Sun that we are led to the following conclusion: scaling from the characteristics of Type II radio bursts associated with solar CME's to stellar flare-related CME's cannot be performed reliably.

**We stress that our interest in the present paper is narrowly focused on one particular aspect of CME's: namely, how CME's are related to the generation of Type II radio bursts. Apart from that topic, we have made no attempt to address how *other* properties of solar CME (e.g. speeds, masses) might be scalable to the corresponding stellar CME properties (as discussed, e.g. by Aarnio et al 2012).**

### Acknowledgements

The authors thank an anonymous referee for an extended report which helped to improve the presentation of the paper. We especially appreciate the referee's suggestion about the possibility of Type II emission in the distant outer wind of a flare star (Section 6).

### REFERENCES

Aarnio, A. N., Stassun, K. G., Hughes, W. J., & MacGregor, S. L. 2011, Solar Phys. 268, 195

Aarnio, A. N., Matt, S. P., & Stassun, K. G. 2012, ApJ 760, 9

Al-Hamadani, F., Pohjolainen, S., & Valtonen, E. 2017, Solar Phys 292, 127

Alvarado-Gomez, J., Drake, J. J., Cohen, O., Moschou, S. P., & Garraffo, C. 2018, ApJ 862, 93

Berdyugina, S. V., Harrington, D. M., Kuzmychov, O. et al. ApJ 847, 61.

Gershberg, R. E. 1970, Flares on Red Dwarf Stars (Moscow: Nauka)

Gopalswamy, N. 2006, J. Astrophys. Astron. 27, 243

Gopalswamy, N., Thompson, W. T., Davila, J. M. et al. 2009, Solar Phys. 259, 227

Habbal, S. R., Druckmuller, M., Morgan, H. et al. 2010, ApJ 719, 1362

Houdebine, E. R., Foing, B. H., & Rodono, M. 1990, A&A 238, 249

Inoue, S., Kusana, K., Buechner, J. & Skala, J. 2018, Nature Commun. 9, 174

Johnstone, C. P. & Guedel, M. 2015, A&A 578, A129

Katsova, M. M., Badalyan, O. G., & Lifshits, M. A. 1987, Soviet Astron. 31, 652

Klassen, A., Aurass, H., Mann, G., & Thompson, B. J. 2000, Astron. Ap. Suppl. 141, 357

Korhonen, H., Vida, K., Leitzinger, M. et al. 2017, in IAU Symposium S328, Living Around Active Stars, edited by Nandy, D., Valio, A., & Petit, P., (Cambridge, England: Cambridge University Press), p. 198

Koskela, J. S., Virtanen, I. I., & Mursula, K. 2017, ApJ 835, 63

Leitzinger, M., Odert, P., Greimel, R. et al 2014, MNRAS 443, 898

MacDonald, J. & Mullan, D. J. 2017, ApJ 850, 58

MacDonald, J., Mullan, D. J., & Dieterich, S. 2018, ApJ 860, 15

MacQueen, R. M., Eddy, J. A., Gosling, J. T. et al. 1974, ApJ Lett. 187, L85

Mann, G., Jansen, F., MacDowall, R. J., Kaiser, M., & Stone, R. G. 1999, Astron. Ap. 348, 614

Mann, G., Klassen, A., Aurass, H., & Classen, H.-T. 2003, Astron. Ap. 400, 329

Moreton, G. 1960, Astron. J. 65, 494

Morin, J., Donati, J.-F., Petit, P. et al. 2008, MNRAS 390, 567

Mostl, C., Amla, K., Hall, J. R. et al. 2014, ApJ 787, 119

Mullan, D. J. 1976, Irish Astron. J. 12, 277

Mullan, D. J. 1979, ApJ 231, 152

Mullan, D. J. 2009, Physics of the Sun: A First Course (Boca Raton: CRC Press), Section 17.14.

Mullan, D. J. & MacDonald, J. 2001, ApJ 559, 353

Mullan, D. J., MacDonald, J., Dieterich, S., & Fausey, H. 2018, ApJ 869, 149

Mullan, D. J., MacDonald, J. & Townsend, R. H. D. 2007, ApJ 670, 1420

Mullan, D. J. & Paudel, R. R. 2018, ApJ 854. 14

Mullan, D. J., Smith, C. W., Ness, N. F., & Skoug, R. M. 2003, ApJ 583, 496

Newkirk, G. A. 1961, ApJ 133, 983

Parker, E. N. 1963, Interplanetary Dynamical Processes (New York: Interscience Publishers)

Schatten, K. H., Wilcox, J. M., & Ness, N. F. 1969, Solar Phys. 6, 442

Schmitt, J. H. M. M., Lemen, J. R., & Zarro, D. 1989, Solar Phys. 121, 361

Smerd, S. F., Sheridan, K. V., & Stewart, R. T. 1975, Ap. Letters 16, 235

Smith, C. W., Mullan, D. J., Ness, N. F., Skoug, R. M., & Steinberg, J. 2001, J. Geophys. Res. 106, 18625

Spitzer, L. 1962, Physics of Fully Ionized Gases, 2$^{nd}$ ed., Interscience, New York.

Shulyak, D., Reiners, A., Engeln, A. et al. 2017, Nature Astron., 1, 184

Tousey, R. 1973, in Proc. of Space Research XIII, vol. 2, p. 713

Uchida, Y. 1968, Solar Phys. 4, 30

Vida, K., Kriscovics, L., Olah, K. et al. 2016, Astron. & Ap 590, 11

Vida, K., Leitzinger, M., Kriscovics, L. et al 2019, arXiv 1901.04229

Warmuth, A., & Mann, G. 2005, Astron. Ap. 435, 1123

Wild, J. P., Smerd, S. F., & Weiss, A. A. 1963, Ann. Rev. Astron. Ap. 1, 291

Wood, B. E. 2018, IOP Conf. Series: Journal of Physics: Conf. Ser. **1100** 012028; doi: 10.1088/1742-6596/100/012028

Yashiro, S., Gopalswamy, N., Michalik, G. et al. 2004, J. Geophys. Res. 109, A07105

Youssef, M. 2012, J. Astron. Geophys. 1, 172

Zaitsev, V. V. 1969, Soviet AJ 12, 610